\documentclass[twocolumn,floats,prl,aps]{revtex4}

\usepackage[dvips]{graphicx}
\usepackage{amsfonts}
\usepackage{amsmath}
\usepackage{amssymb}
\usepackage{exscale}
\usepackage{eufrak}

\begin{document}

\draft    


\title{Kinetic energy change with doping upon superfluid condensation 
	   in high temperature superconductors
      }

\author{Guy Deutscher$^{1}$, Andr\'es Felipe Santander-Syro$^{2}$, and Nicole Bontemps$^{3}$}

\address{$^{1}$School of Physics and Astronomy, 
		 Tel Aviv University, Ramat Aviv 69978, Isra\"el}

\address{$^{2}$Laboratoire de Physique des Solides, CNRS UMR 8502,
         Universit\'e Paris-Sud, 91405 Orsay cedex, France}
         
\address{$^{3}$Laboratoire de Physique du Solide, CNRS UPR 5,
         Ecole Sup\'erieure de Physique et Chimie
         Industrielles de la Ville de Paris,
         75231 Paris cedex 5, France}

\date{\today}


\begin{abstract}
In conventional BCS superconductors, the electronic kinetic energy increases 
upon superfluid condensation (the change $\Delta E_{kin}$ is positive). 
Here we show that in the high critical temperature superconductor 
Bi$_2$Sr$_2$CaCu$_2$O$_{8+\delta}$, $\Delta E_{kin}$ crosses over 
from a fully compatible conventional BCS behavior ($\Delta E_{kin}>0$) 
to an unconventional behavior ($\Delta E_{kin}<0$) 
as the free carrier density decreases. If a single mechanism is responsible 
for superconductivity across the whole phase diagram 
of high critical temperature superconductors, this mechanism should allow 
for a smooth transition between such two regimes around optimal doping.

\end{abstract}


\maketitle  



One of the fundamental predictions of the BCS theory is that the kinetic energy 
of the charge carriers increases upon condensation in the superconducting state, 
while the interaction energy decreases and overcompensates the kinetic energy increase, 
resulting in a net energy gain. The value of this condensation energy is 
easily determined, for instance from the value of the thermodynamical critical field, 
but the respective changes in kinetic and interaction terms are not easily accessed. 
In fact, the change in kinetic energy in ``conventional" BCS superconductors 
has never been determined experimentally. This change is of the order of
$(\Delta/E_{F})^{2}$, where $\Delta$ is the energy gap and $E_{F}$ 
the Fermi energy. It is exceedingly small for a typical low temperature superconductor, 
of the order of $10^{-6}$ to $10^{-8}$.

The situation is much more favorable in High Critical Temperature Superconductors 
(HCTS, cuprates), where the gap is larger and the Fermi energy smaller, 
so that the change in kinetic energy, if it is conform to the predictions 
of the BCS theory, should be of the order of $10^{-3}$ to $10^{-2}$, 
a change that has become accessible experimentally~\cite{R1,R2,R3}. 
However, the mechanism for HCTS is still under debate and the change 
in kinetic energy could well be different from that predicted by BCS, 
including in sign.

It is of particular interest to investigate the case of overdoped high temperature 
superconductors. There is a general belief that in the overdoped range, 
the cuprates can be described in their normal state as Fermi liquids. 
Thus it is conceivable that in this regime, condensation is of the BCS kind. 
And if it is, according to the above considerations regarding orders of magnitude, 
the change in kinetic energy should be large enough to be measured, 
allowing a quantitative comparison between theory and experiment.

Our analysis shows that the change in kinetic energy in overdoped 
Bi$_2$Sr$_2$CaCu$_2$O$_{8+\delta}$ (Bi-2212) having $T_{c}=63$~K 
is indeed compatible with the predictions of the BCS theory, 
both in sign and in size. The latter result appears in the data in our previous papers, 
but was not explicitly mentioned~\cite{R2,R3}.  This is in contrast with the change 
of kinetic energy in optimally doped, and definitely in underdoped Bi-2212, 
which has been found to be of the opposite sign~\cite{R1,R2,R3}. We observe that going 
from the overdoped to the underdoped regime, the change in kinetic energy 
is actually progressive, going through zero not far from optimum doping. 
This progressive change strongly suggests that there is in the cuprates 
a smooth transition from a conventional mode of condensation in the overdoped regime 
to an unconventional mode in the underdoped one. 

We recall that from measurements of the reflectivity, one can derive the real part
$\sigma_{1}(\omega)$ and the imaginary part $\sigma_{2}(\omega)$ 
of the optical conductivity.  The single band sum rule~\cite{R4} writes:
\begin{equation}
	\int_{0}^{\infty} \sigma_{1,xx}(\omega) d\omega = \frac{\pi e^{2} a^{2}}
						{2 \hbar^{2} V} E_{K},
\label{EqSumRule}
\end{equation}
where $e$ is the electron charge, $a$ the in-plane lattice constant, $V$ the volume 
of the unit cell. $E_{K}$ is given by:
\begin{equation}
	E_{K}=\frac{2}{a^{2} N} \sum_{k}\frac{\partial^{2}\epsilon_{k}}{\partial k_{x}^{2}}
				n_{k},
\label{EqEK}
\end{equation}
where $N$ is the number of $k$ vectors, $\epsilon_{k}$ is the dispersion 
from the kinetic energy part of the hamiltonian, and $n_{k}$ is the 
momentum distribution function.
 
In a nearest neighbor tight binding model, the kinetic energy is related to $E_{K}$:
\begin{equation}
	E_{kin} = -E_{K}.	
\label{EqEkin}
\end{equation}

It was argued that relation~(\ref{EqEkin}) is still valid (however within $\sim 50$\%) 
when taking into account the second nearest neighbor hopping~\cite{R5}.

The optical conductivity is generally derived experimentally from a 
Kramers-Kronig transform of the reflectivity~\cite{R6}, 
or by fitting the reflectivity~\cite{R2,R3}
or more accurately by a combination of ellipsometric measurements in the visible 
supplemented by infrared and visible reflectivity~\cite{R1}. 
Two difficulties arise when computing the spectral weight defined in Eq.\ref{EqSumRule}, 
related to the limits of the integral: {\it i)} one has to choose a cut-off frequency  
$\Omega_{c}$ in order to avoid including interband transitions which are irrelevant 
to the calculation of the kinetic energy, {\it ii)} the optical conductivity 
cannot be derived starting from zero: one is restricted at best to the experimental 
lowest frequency. 

The choice of the high frequency cut-off is a difficult problem: it is generally agreed 
to select a cut-off significantly below the energy of the charge transfer band, 
located at typically 1.5~eV. Therefore according to the authors,  
$\Omega_{c} \sim 0.6-1.2$~eV \cite{R1,R2,R3,R7}. We have calculated the spectral weight 
for various cut-off energies in this range.  It changes with the cut-off 
but the trend as a function of doping is robust. In the following, we will show data 
for a 1~eV cut-off. The uncertainty on the spectral weight is $\sim 0.3$\% 
in this range~\cite{R8}.
The low limit can be dealt with by fitting the reflectivity and using 
the deduced optical conductivity in order to extrapolate to zero~\cite{R2,R3}. 
In the superconducting state, the spectral weight includes the superfluid weight 
which is extracted from the data [9] or inferred from the fit~\cite{R2,R3}.

\begin{figure}
  \begin{center}
    \includegraphics[width=8cm]{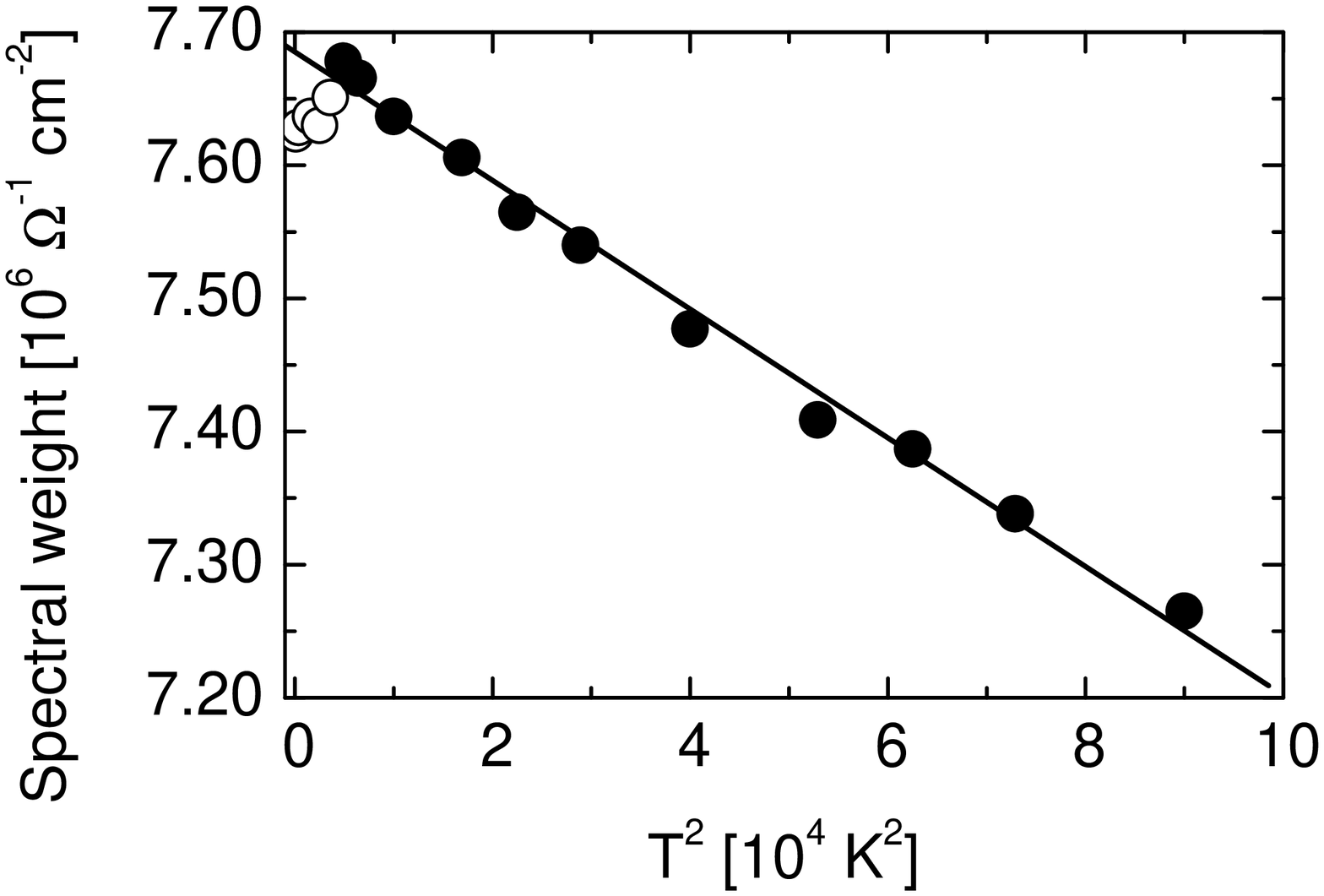}
  \end{center}
  \caption{\label{Fig1}
         Spectral weight of the overdoped Bi-2212 sample, integrated up to 1~eV, 
         plotted versus $T^2$, from Ref.\cite{R3}. 
         Full symbols: spectral weight in the normal state, 
         open symbols: spectral weight in the superconducting state, 
         including the weight of the superfluid.
         }
\end{figure}

We show in figure~\ref{Fig1} the change of spectral weight 
as a function of  $T^{2}$, for an overdoped 
Bi-2212 sample, after reconstructing the optical conductivity through a well 
controlled fitting procedure from the reflectivity of a thin film~\cite{R3,R8}. 
The spectral weight integral (Eq.\ref{EqSumRule}) has been extended up to 1~eV
(8000~cm$^{-1}$). In the normal state, it is linear in $T^{2}$. 
The change from room temperature down to $T_{c}$ is of about 5\%.  
At $T_{c}$, a change in {\it sign} of the slope of the temperature dependence of 
the integral, corresponding to an increase in kinetic energy, is observed very clearly. 
By extrapolating the temperature dependence in the normal state down to $T=0$ 
following the $T^{2}$ behavior found above $T_{c}$, one can obtain the value 
of the difference between the kinetic energy in the normal and superconducting states 
in that limit. We find that it is of about 1\%. 

According to BCS theory, the increase in kinetic energy in the superconducting state 
per unit volume is given by:
\begin{equation}
	\Delta E_{kin}=\frac{\Delta^{2}}{\mathcal{V}} - \frac{\mathcal{N}(0)\Delta^{2}}{2},
\label{EqDEkBCS-Exact}
\end{equation}
where $\mathcal{N}(0)$ is the density of states at the Fermi level, and 
$\mathcal{V}$ the interaction parameter. Because $\mathcal{N}(0)\mathcal{V}$ 
is in any case substantially smaller than unity, we neglect at first 
the second term of the right hand side. We then obtain:
\begin{equation}
	\frac{\Delta E_{kin}}{E_{kin}^{N}} \approx \frac{1}{\mathcal{N}(0)\mathcal{V}}
				\left(\frac{\Delta}{E_{F}}\right)^2,
\label{EqDEkBCS-Approx}
\end{equation}
where $E_{kin}^{N}$ is the kinetic energy in the normal state at $T = 0$. 
The quantity $\Delta E_{kin}/E_{kin}^{N}$ is precisely that measured experimentally 
following the procedure described above. Taking the values  
$\Delta=20$~meV and $E_{F}=500$~meV \cite{R10}, we obtain 
$\mathcal{N}(0)\mathcal{V} \approx 0.16$, not an unreasonable value. 
A slightly higher value of $0.2$ is obtained by taking into account 
the condensation energy, {\it i.e} the second term of the right hand side of 
Eq.\ref{EqDEkBCS-Exact}, as measured for instance by Loram~\cite{R11}. 
The measured change in kinetic energy for this overdoped sample is thus 
in good agreement with BCS theory, both in sign and in value.

\begin{figure}
  \begin{center}
    \includegraphics[width=8cm]{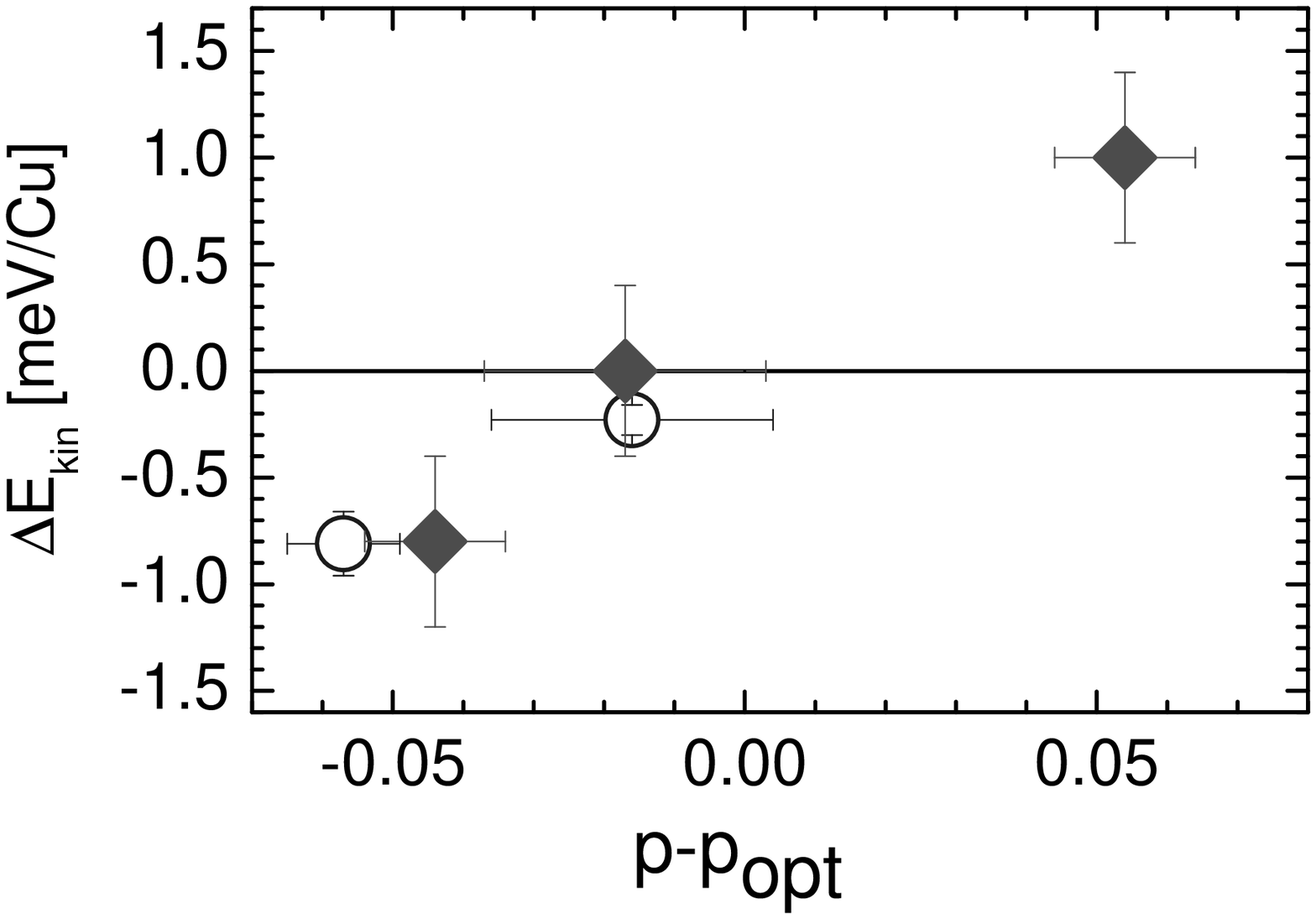}
  \end{center}
  \caption{\label{Fig2}
         Change $\Delta E_{kin}$ of the kinetic energy, in meV per copper site,
         calculated from equations (\ref{EqSumRule}) and (\ref{EqEkin}), 
         versus the charge $p$ per copper with respect to $p_{opt}$ (Eq.\ref{EqTallon}). 
         Full diamonds: data from Ref.\cite{R3}, high frequency cut-off 1~eV. 
         Open circles: data from Ref.\cite{R1},  high frequency cut-off 1.25~eV.
         Error bars: vertical, uncertainties due to the extrapolation of the 
         temperature dependence of the normal state spectral weight 
         down to zero temperature; 
         horizontal, uncertainties resulting from $T_{c}/T_{c,max}$ 
         through Eq.\ref{EqTallon} (see text). We have taken 
         $T_{c,max}=(83 \pm 2)$~K for films and $(91 \pm 2)$~K for crystals.
         }
\end{figure}

We now turn to a comparison between the behavior of the overdoped sample, 
and that of optimally and underdoped samples. In the normal state, the change 
in kinetic energy with temperature is somewhat smaller (4\%) 
but close to that of the overdoped sample~\cite{R1,R2,R3}. However, upon
condensation, there is now a relative decrease in kinetic energy 
of about $-0.2$\%  for close to optimally doped~\cite{R1}
or about 0 within the error bars~\cite{R2,R3}, and of $-0.5$\% \cite{R1,R12}
or $-0.7$\% \cite{R3,R13} for underdoped samples. 
The transition between the BCS and unconventional regimes 
thus appears to be progressive: Figure~\ref{Fig2} shows the change $\Delta E_{kin}$
(in meV/Cu) as a function of $(p-p_{opt})$ \cite{R14}, through 
\begin{equation}
	 \frac{T_{c}}{T_{c,opt}} = 1-86.2(p-p_{opt})^{2},
\label{EqTallon}
\end{equation}
where $p$ is the charge per Cu atom, and $p_{opt}$ corresponds to the maximal 
critical temperature $T_{c,opt}$ \cite{R15}. 

The case of the overdoped sample is clear. The kinetic energy increases in the 
superconducting state by an amount compatible with a BCS condensation. 
This result is in line with the observation that the full spectral weight 
in the superconducting state is recovered, within $\Delta E_{kin}$, at an energy 
equal to a few times the gap, as shown in~\cite{R2,R3}. 
By contrast, for the underdoped sample, 
the full spectral weight is clearly only recovered at energies of more than 1~eV, 
or about 40 times the gap, and the kinetic energy now decreases in the 
superconducting state \cite{R2,R3}. Both in terms of the change in kinetic energy and 
rate of recovery of the spectral weight, nearly optimum doped samples 
are intermediate between the overdoped and the underdoped ones:  
the change in kinetic energy is small.  This doping dependence suggests 
a smooth transition from a BCS mode of condensation in the overdoped regime 
to a different mode in underdoped samples, as one would expect for instance 
in the case of a BCS to Bose-Einstein crossover~\cite{R16}.

In the case of YBa$_2$Cu$_3$O$_{6+x}$ (YBCO), we are not aware of any measurement 
in overdoped samples. In underdoped YBa$_2$Cu$_3$O$_{6.6}$, an unconventional 
energy scale (of about 0.6~eV) for recovering the full spectral weight 
was found~\cite{R17}. In optimally doped samples, the change in kinetic energy 
is definitely small~\cite{R17,R18}.

While much theoretical and experimental emphasis has been given in previous works 
to the unconventional behavior of the kinetic energy change $\Delta E_{kin}$ 
in underdoped samples, the full compatibility of the behavior of overdoped samples 
with a BCS mode of condensation has been so far overlooked. Most important, 
the sign and size of $\Delta E_{kin}$ upon condensation in the superconducting state 
and the rate of recovery of the spectral weight point simultaneously towards 
a progressive change in the condensation regime when going from underdoped to overdoped. 
Such an overall behavior shows that the high-$T_c$ mechanism, if it is the same 
across the phase diagram, must allow for the observed transition 
from kinetic energy loss to kinetic energy increase as doping is increased.

\acknowledgments

One of us (G.D.) wishes to acknowledge the support of ESPCI through 
Chaire Paris-Sciences during the course of this work, 
as well as the support of the Oren Family Chair of Experimental Solid State Physics.



\begin{references}
%
\bibitem{R1} H. J. A. Molegraaf {\it et al.}, Science {\bf 295}, 2239 (2002).
\bibitem{R2} A.F. Santander-Syro {\it et al.}, Europhys.~Lett {\bf 62}, 568 (2003).
\bibitem{R3} A.F. Santander-Syro {\it et al.}, 
		Phys.~Rev.~B {\bf 70}, 134504 (2004).
\bibitem{R4} M. Norman and C. P\'epin, Rep.~Prog.~Phys.~{\bf 66}, 1547 (2003).
\bibitem{R5} D. van der Marel {\it et al.}, condmat/0302169 (2003).
\bibitem{R6} D. Tanner and T. Timusk, { \it Optical properties 
		of high temperature superconductors}, in ``Physical Properties of High Critical 
		Temperature Superconductors III", Ed. D. Ginsberg , World Scientific, 
		Singapore (1992).
\bibitem{R7} A. J. Millis {\it et al.}, condmat/0411172 (2004).
\bibitem{R8} A. Zimmers, PhD thesis (unpublished).
\bibitem{R9} S. Dordevic {\it et al.}, Phys.~Rev.~B {\bf 65}, 134511 (2002).
\bibitem{R10} M. R. Norman {\it et al.}, Phys.~Rev.~B {\bf 52}, 615 (1995).
\bibitem{R11} J. W. Loram {\it et al.}, Physica C {\bf 341-348}, 831 (2001).
\bibitem{R12} Reference~\cite{R1} displays data on an ``optimally doped" 
		Bi-2212 single crystal with $T_{c}=88$~K, whereas at optimal doping, 
		Bi-2212 crystals easily achieve $T_{c}=90$~K. We have assumed here that 
		this particular sample is slightly underdoped.
\bibitem{R13} The normal state spectral weight follows a $T^2$ dependence 
		from room temperature down to 100~K in the underdoped sample of 
		Ref.\cite{R1}, whereas its value saturates below 150~K 
		in Ref.\cite{R3}. We extrapolated the normal state spectral weight 
		down to $T=0$ using the $T^2$ behavior in Ref.\cite{R1}
		and the saturated value in Ref.\cite{R3}.
\bibitem{R14} A similar graph was shown in M.~Norman and C.~P\'epin, 
		Phys.~Rev.~B {\bf 66}, 100506(R) (2002), 
		where $E_{kin}$ has been calculated versus doping in a specific model. 
		However, our experimental data shown in this graph were taken from 
		Ref.\cite{R2}, where the positive sign of $\Delta E_{kin}$ 
		in the overdoped sample was overlooked.
\bibitem{R15} J. L. Tallon and J. W. Loram, 
		Physica C {\bf 349}, 53 (2001).
\bibitem{R16} P. Nozi\`eres and S. Schmitt-Rink, 
		J.~Low.~Temp.~Physics {\bf 59}, 195 (1985); 
		A. Leggett, in {\it Modern Trends in the theory of condensed matter},
		Ed. A Bekalksky and J. Przyspawa (Springer Verlag, 1980), pp.13-27.
\bibitem{R17} C. C. Homes {\it et al.}, Phys.~Rev.~B~{\bf 69}, 024514 (2004).
\bibitem{R18} The sign of  $\Delta E_{kin}$ in underdoped and close to optimally doped 
		Bi-2212 (Ref.\cite{R1,R2,R3}) has been challenged 
		by Boris {\it et al.}, Science {\bf 304}, 708 (2004). 
		These authors present a quantitative analysis for optimally doped YBCO 
		and claim a similar behaviour for slightly underdoped Bi-2212.  
		Their analysis and the resulting conclusions are objected by D.~van der Marel 
		(private communication), A.F. Santander-Syro and N. Bontemps.

\end{references}
\end{document}